\documentclass[twocolumn,aps,prl,showpacs]{revtex4}
\usepackage{graphics}
\begin{document}

\title{Colloidal glass transition: beyond mode-coupling theory}
\author{Grzegorz Szamel}
\affiliation{Department of Chemistry, 
Colorado State University, Fort Collins, CO 80525}

\date{\today}


\begin{abstract}
A new theory for dynamics of concentrated colloidal suspensions 
and the colloidal glass transition is proposed. The starting point
is the memory function representation of the 
density correlation function. The memory function can be expressed 
in terms of a time-dependent pair-density correlation 
function. An exact, formal equation of motion for this 
function is derived and a factorization approximation is applied  
to its evolution operator. In this way a closed set of equations 
for the density correlation function and the memory function is obtained. 
The theory predicts an ergodicity breaking
transition similar to that predicted by the mode-coupling
theory, but at a higher density.  
\end{abstract}
\maketitle

There has been a lot of interest in recent years in the theoretical
description of dynamics of concentrated 
suspensions and the colloidal glass transition \cite{review}. 
It has been stimulated by ingenious experiments
which provide detailed information about microscopic dynamics
of colloidal particles \cite{expts}. Due to the abundance of experimental data 
the colloidal glass transition has emerged 
as a favorite, model glass transition to be studied \cite{colloglass}.

One of the conclusions of these studies
is the acceptance of the mode-coupling theory (MCT)
as the theory for dynamics of concentrated suspensions and their 
glass transition \cite{Cates}. Historically, this is somewhat surprising since 
MCT was first formulated for simple fluids with Newtonian
dynamics \cite{Goetze} and only afterwards was adapted to colloidal systems
with stochastic (Brownian) dynamics \cite{SL}. On the other hand, 
basic approximations of MCT are 
less severe for Brownian systems \cite{commentHI}.

MCT is a theory for correlation functions of 
slow variables, \emph{i.e.} variables satisfying local conservation laws.
For Brownian systems there is only one such variable: local density.
MCT's starting point is the memory function representation
of the density correlation function \cite{commentMCTder,commentirr}.
The memory function is 
expressed in terms of a time-dependent pair-density (\emph{i.e.}
four-particle) correlation
function evolving with so-called projected dynamics. For Brownian
systems this step is exact \cite{commentBvN}. 
The central approximation of MCT is the factorization approximation
in which the pair-density correlation function is replaced by a product 
of two time-dependent density correlation functions.
As a result one obtains a closed, nonlinear equation of motion
for the density correlation function. This equation predicts an
ergodicity breaking transition that is identified with the
colloidal glass transition.
MCT has also been used to describe, \emph{e.g.}, 
linear viscoelasticity \cite{Naegele}, 
dynamics of sheared suspensions \cite{Fuchs}, 
and colloidal gelation \cite{DawsonFuchs}. 
By and large, its predictions agree with experimental and
simulational results \cite{Cates,vanM}.

In spite of these successes, MCT's problems are well known \cite{Cates}.
The most important, fundamental problem
is that once the factorization approximation is made there is 
no obvious way to extended and/or improve the theory.
This is most acute for Brownian systems because there
the density is the only slow mode and thus couplings to other modes
cannot be invoked! Furthermore, MCT systematically 
overestimates so-called dynamic feedback effect. Thus, \emph{e.g.},
it underestimates
the glass transition volume fraction for a Brownian hard-sphere system
(by about 10\% \cite{Cates}) 
and overestimates the glass transition temperature
for a Lennard-Jones mixture (by a factor of 2 \cite{Kob}).
Finally, MCT cannot describe slow dynamics in systems 
without static correlations \cite{SS}.

A way to improve upon MCT would be to introduce
many-particle dynamic variables into the theory. Such an 
attempt has been made for simple fluids \cite{Oppen}; it was argued that
these variables (essentially, pair-density fluctuations) describe
clusters of correlated particles. Unfortunately no quantitative results
have been reported based on this interesting approach.

We propose a different way to go beyond MCT. 
Rather than factorizing the pair-density correlation
function, we derive an exact, formal equation of motion for it \cite{Arun}. 
The structure of this equation is very similar to that of the 
equation of motion for the density correlation function; 
``pair'' analogues of the usual frequency matrix and the irreducible 
memory function can be identified. The basic approximation
of our theory is a factorization of the evolution operator
of the pair-density correlation function. 
After this approximation
we obtain a closed system of equations of motion for the
density correlation function and the memory function. These equations
predict an ergodicity breaking transition; for a Brownian hard sphere
system the glass transition volume fraction, $\phi_g$, is equal to $.549$
(note that $\phi_{g}^{MCT} = .525$, $\phi_{g}^{exp} \approx .58$).

Our theory is similar to MCT in that it relies upon an uncontrollable
factorization approximation. In contrast to MCT, it 
uses this approximation one step later. Thus, \textit{e.g.}, our theory preserves 
the memory function representation of the pair-density correlation
function while MCT approximates the latter by a product of two
density correlation functions. However, as usual in the liquid
state theory, \textit{a priori} these features do not guarantee the
superiority of our approach as compared to MCT.

Our theory starts from the memory
function representation of the density correlation function, $F(k;t)$, 
\begin{equation}\label{Fktdef}
F(k;t) = \frac{1}{N} \left< n(\mathbf{k}) \exp(\Omega t)n(-\mathbf{k})
\right>.
\end{equation}
Here $N$ is the number of particles, 
$n(\mathbf{k})$ is the Fourier transform of the density,
$
n(\mathbf{k}) = \sum_l e^{-i\mathbf{k}\cdot\mathbf{r}_l},
$
and $\Omega$ is the $N$-particle evolution operator, 
\textit{i.e.} the Smoluchowski operator,
$
\Omega = D_0 \sum_l \frac{\partial}{\partial\mathbf{r}_l}\cdot
\left(\frac{\partial}{\partial\mathbf{r}_l}-\beta\mathbf{F}_l\right)
$
\cite{commentHI2},
with $D_0$ being the diffusion coefficient of an isolated Brownian
particle, $\beta=1/(k_B T)$, and $\mathbf{F}_l$ a force acting
on particle $l$. 
Finally, $\left<\dots\right>$ denotes
the canonical ensemble average; the equilibrium distribution
stands to the right of the quantity being averaged, and all operators
act on it as well as on everything else. Usually, the memory function
representation of the Laplace transform of the density correlation
function, $F(k;z)$, is written as \cite{CHess}
\begin{equation}\label{Mkt}
F(k;z)=\frac{S(k)}{z+\frac{D_0k^2}{S(k)(1+M(k;z))}}
\end{equation}
where $S(k)$ is the static structure factor and 
$M(k;z)$ is the Laplace transform of the irreducible memory function.
We re-write (\ref{Mkt}) in a form that 
will allow us to identify the pair analogues
of the frequency matrix and the memory function. We write
a memory function expression for the Laplace transform ($LT$)
of $\dot{F}(k;t)$
\begin{eqnarray}\label{Fder}
LT(\dot{F}(k;t)) &=&   - \mathbf{k} \cdot 
\left(\mathsf{1} + \mathsf{M}(\mathbf{k};z)\mathsf{O}^{-1} 
\right)^{-1} \mathsf{O} \cdot\mathbf{k} \; \nonumber \\
&& \times \frac{1}{\left<n(\mathbf{k}) n(-\mathbf{k})\right>}   \;F(k;z).
\end{eqnarray}
Here $\mathsf{1}$ denotes a unit 3d tensor, $\mathsf{O}$ is defined through
$
- \mathbf{k}\cdot\mathsf{O}\cdot\mathbf{k} = 
\left<n(\mathbf{k})\Omega n(-\mathbf{k})\right>
$ 
(note that $\mathsf{O} = \mathsf{1} D_0 N$),
and $\mathsf{M}(\mathbf{k};z$) is the Laplace transform of the 
current correlation function evolving with
projected dynamics,
\begin{equation}\label{mktdef}
\mathsf{M}(\mathbf{k};t) = 
\left< \mathbf{j}(\mathbf{k}) \exp(\Omega^{irr}t) \mathbf{j}(-\mathbf{k})
\right>.
\end{equation}
where $\mathbf{j}(\mathbf{k})$ is a projected current density,
\begin{equation}\label{current}
\mathbf{j}(\mathbf{k}) = \hat{Q}_n D_0 \sum_l  
(-i\mathbf{k} + \beta\mathbf{F}_l)e^{-i\mathbf{k}\cdot\mathbf{r}_l}. 
\end{equation}
In Eq. (\ref{current}) 
$\hat{Q}_n=1-\hat{P}_n$, and $\hat{P}_n$ is a projection operator 
on the density subspace,
\begin{equation}
\hat{P}_n = \dots 
\sum_{\mathbf{q}}
n(-\mathbf{q})\left>\frac{1}{\left<n(\mathbf{q})n(-\mathbf{q})\right>}
\right<n(\mathbf{q})\dots .
\end{equation}
Finally, in Eq. (\ref{mktdef}) 
$\Omega^{irr}$ is the ``one-particle irreducible Smoluchowski
operator'' \cite{CHess},
\begin{equation}
\Omega^{irr} = \hat{Q}_n \sum_l \frac{\partial}{\partial\mathbf{r}_l}
\hat{Q}_l \cdot
\left(\frac{\partial}{\partial\mathbf{r}_l}-\beta\mathbf{F}_l\right)
\hat{Q}_n,
\end{equation}
where $\hat{Q}_l = 1 - \hat{P}_l$, and the projection operator $\hat{P}_l$
reads 
\begin{equation}
\hat{P}_l = \dots \sum_{\mathbf{q}}e^{i\mathbf{q}\cdot\mathbf{r}_l}\left>
\right<e^{-i\mathbf{q}\cdot\mathbf{r}_l} \dots .
\end{equation}
To make connection with 
the usual form of the memory function representation 
we note that 
$
\mathbf{k}\cdot\mathsf{O}\cdot\mathbf{k}
/\left<n(\mathbf{k})n(-\mathbf{k})\right> = D_0 k^2/S(k)
$ 
is the frequency matrix
and 
$
\hat{\mathbf{k}}\cdot\mathsf{M}(\mathbf{k};z)
\mathsf{O}^{-1}\cdot\hat{\mathbf{k}} = M(k;z),
$ 
where $\hat{\mathbf{k}}=\mathbf{k}/k$, is the irreducible memory function.

To obtain a convenient expression for $\mathsf{M}(\mathbf{k};t)$
in terms of a pair-density correlation function we use the following
exact \cite{comment2b} equality:
\begin{eqnarray}\label{projection}
\mathbf{j}(-\mathbf{k}) &=& 
\sum_{(\mathbf{k}_1,\mathbf{k}_2)} \sum_{(\mathbf{k}_3,\mathbf{k}_4)} 
n_2(-\mathbf{k}_1,-\mathbf{k}_2) \nonumber \\ &&\times
g(\mathbf{k}_1,\mathbf{k}_2;\mathbf{k}_3,\mathbf{k}_4)
\left<n_2(\mathbf{k}_3,\mathbf{k}_4)\mathbf{j}(-\mathbf{k})\right>.
\end{eqnarray}
Here $n_2(\mathbf{k}_1,\mathbf{k}_2)$ is the part of pair-density
fluctuations orthogonal to the one-particle density fluctuations,
\begin{equation}
n_2(\mathbf{k}_1,\mathbf{k}_2) = 
\hat{Q}_n \sum_{l\neq m} e^{-i\mathbf{k}_1\cdot\mathbf{r}_l
- i\mathbf{k}_2\cdot\mathbf{r}_m}.
\end{equation}
Furthermore, in Eq. (\ref{projection}) the sums over
$\mathbf{k}_i < \mathbf{k}_{i+1}$ are understood and 
$g$ denotes
the inverse pair-density fluctuations matrix (it is a pair analogue of 
$1/\left<n(\mathbf{k})n(-\mathbf{k})\right>$),
\begin{eqnarray}
&&\sum_{(\mathbf{k}_3,\mathbf{k}_4)}
g(\mathbf{k}_1,\mathbf{k}_2;\mathbf{k}_3,\mathbf{k}_4)
\left<n_2(\mathbf{k}_3,\mathbf{k}_4)n_2(-\mathbf{k}_5,-\mathbf{k}_6)\right>=
\nonumber \\
&&\delta_{\mathbf{k}_1,\mathbf{k}_5}\delta_{\mathbf{k}_2,\mathbf{k}_6}.
\end{eqnarray}

Using identity (\ref{projection}) we can express memory function 
(\ref{mktdef}) in terms
of the time-dependent pair-density correlation function
evolving with one-particle irreducible dynamics,
\begin{eqnarray}
&& F_{22}(\mathbf{k}_1,\mathbf{k}_2;\mathbf{k}_3,\mathbf{k}_4;t)= 
\nonumber \\
&& \left<n_2(\mathbf{k}_1,\mathbf{k}_2) \exp(\Omega^{irr}t)
n_2(-\mathbf{k}_3,-\mathbf{k}_4)\right>.
\end{eqnarray}
Rather than factorizing $F_{22}$, we use the projection operator
method to derive an exact, formal
equation of motion for this function. The derivation  
will be given elsewhere \cite{GS}; here we present the structure
of the final formula for the Laplace transform of the
time-derivative of the pair-density correlation function, 
$\dot{F}_{22}$,
\begin{widetext}
\begin{eqnarray}\label{F22der}
LT\left(
\dot{F}_{22}(\mathbf{k}_1,\mathbf{k}_2;\mathbf{k}_3,\mathbf{k}_4;t)\right) 
&=-& 
\displaystyle{\prod_{i=3}^{7} 
\left(\sum_{(\mathbf{k}_{2i-1},\mathbf{k}_{2i})}\right)}
\left(\mathbf{k}_1,\mathbf{k}_2\right)
\left(\mathcal{I} +
\mathcal{M}(\mathbf{k}_5,\mathbf{k}_6;\mathbf{k}_7,\mathbf{k}_8;z)
\mathcal{O}(\mathbf{k}_7,\mathbf{k}_8;\mathbf{k}_9,\mathbf{k}_{10})^{-1}
\right)^{-1} \nonumber \\ 
&&
\mathcal{O}(\mathbf{k}_9,\mathbf{k}_{10};\mathbf{k}_{11},\mathbf{k}_{12})
\left(\begin{array}{c}
\mathbf{k}_{11} \\ \mathbf{k}_{12} 
\end{array}\right) 
g(\mathbf{k}_{11},\mathbf{k}_{12};\mathbf{k}_{13},\mathbf{k}_{14})
F_{22}(\mathbf{k}_{13},\mathbf{k}_{14};\mathbf{k}_3,\mathbf{k}_4;z)
\end{eqnarray}
\end{widetext}
In Eq. (\ref{F22der}) $\mathcal{I}$ denotes a unit 6d tensor, 
$\mathcal{O}$ and $\mathcal{M}$ are block matrices, \emph{e.g.}
\begin{eqnarray}
&&\mathcal{O}(\mathbf{k}_1,\mathbf{k}_2;\mathbf{k}_3,\mathbf{k}_4)=
\nonumber \\
&&\left(\begin{array}{cc}
\mathcal{O}_{11}(\mathbf{k}_1,\mathbf{k}_2;\mathbf{k}_3,\mathbf{k}_4) & 
\mathcal{O}_{12}(\mathbf{k}_1,\mathbf{k}_2;\mathbf{k}_3,\mathbf{k}_4)\\
\mathcal{O}_{21}(\mathbf{k}_1,\mathbf{k}_2;\mathbf{k}_3,\mathbf{k}_4) &
\mathcal{O}_{22}(\mathbf{k}_1,\mathbf{k}_2;\mathbf{k}_3,\mathbf{k}_4)
\end{array}\right),
\end{eqnarray}
and the following short-hand notation is used:
\begin{equation}
\left(\mathbf{k}_1,\mathbf{k}_2\right)
\mathcal{O}(\mathbf{k}_1,\mathbf{k}_2;\mathbf{q}_1,\mathbf{q}_2)
\left(\begin{array}{c}
\mathbf{q}_1 \\ \mathbf{q}_2
\end{array}\right) = 
\sum_{i,j} \mathbf{k}_i\cdot\mathcal{O}_{ij}\cdot\mathbf{q}_j.
\end{equation}
$\mathcal{O}$ and $\mathcal{M}$ are the pair analogues of
$\mathsf{O}$ and $\mathsf{M}$ (compare Eqs. (\ref{Fder}) and
(\ref{F22der})); in particular 
\begin{eqnarray}
&-&\left(\mathbf{k}_1,\mathbf{k}_2\right)
\mathcal{O}(\mathbf{k}_1,\mathbf{k}_2;\mathbf{k}_3,\mathbf{k}_4)
\left(\begin{array}{c}
\mathbf{k}_3 \\ \mathbf{k}_4 
\end{array}\right) 
= \nonumber \\
&&\left<n_2(\mathbf{k}_1,\mathbf{k}_2)\;\Omega^{irr}\;
n_2(-\mathbf{k}_3,-\mathbf{k}_4)\right>,
\end{eqnarray}
and $\mathcal{M}_{ij}$ are pair-current correlations 
evolving with a two-particle irreducible evolution
operator $\Omega^{2irr}$, \textit{e.g.},
\begin{eqnarray}
&&\mathcal{M}_{11}(\mathbf{k}_1,\mathbf{k}_2;\mathbf{k}_3,\mathbf{k}_4;t) =
\nonumber \\ 
&& \left<\mathbf{j}_2(\mathbf{k}_1,\mathbf{k}_2)  
\exp(\Omega^{2irr}t) \mathbf{j}_2(-\mathbf{k}_3,-\mathbf{k}_4)
\right>
\end{eqnarray}
where, \textit{e.g.},
\begin{equation}
\left. \mathbf{j}_2(-\mathbf{k}_3,-\mathbf{k}_4)\right> = 
\hat{Q}_{n_2} D_0 \sum_{l\neq m} \frac{\partial}{\partial\mathbf{r}_l}
\hat{Q}_l 
\left.
e^{i\mathbf{k}_3\cdot\mathbf{r}_l+i\mathbf{k}_4\cdot\mathbf{r}_m}
\right>.
\end{equation}
Explicit formulae for $\mathcal{O}$ and $\mathcal{M}$ (including
definitions of $\Omega^{2irr}$ and 
$\hat{Q}_{n_2}$) will be given elsewhere \cite{GS}.

The main approximation of our theory is factorization
of the evolution operator for $F_{22}$. 
Within this approximation the diagonal blocks of 
$\mathcal{O}$ and $\mathcal{M}$ are given by
\begin{eqnarray}\label{factO}
&&\mathcal{O}_{11}(\mathbf{k}_1,\mathbf{k}_2;\mathbf{k}_3,\mathbf{k}_4)=
\mathcal{O}_{22}(\mathbf{k}_2,\mathbf{k}_1;\mathbf{k}_4,\mathbf{k}_3)=
\nonumber \\
&&N \mathsf{O} S(k_2)\delta_{\mathbf{k}_1,\mathbf{k}_3}  
\delta_{\mathbf{k}_2,\mathbf{k}_4}, \\ \label{factm0}
&&\mathcal{M}_{11}(\mathbf{k}_1,\mathbf{k}_2;\mathbf{k}_3,\mathbf{k}_4;t)= 
\mathcal{M}_{22}(\mathbf{k}_2,\mathbf{k}_1;\mathbf{k}_4,\mathbf{k}_3;t)=
\nonumber \\
&&N \mathsf{M}(\mathbf{k}_1;t)F(k_2;t)\delta_{\mathbf{k}_1,\mathbf{k}_3}  
\delta_{\mathbf{k}_2,\mathbf{k}_4},
\end{eqnarray}
and the off-diagonal ones vanish. 
Consistently, we also factorize $g$ and $F_{22}(t=0)$.

Using (\ref{factO}--\ref{factm0}) we can express $F_{22}$
in terms of the density correlation function and the memory function
(note that $F_{22}$ does \emph{not} factorize for $t> 0$).
Substituting $F_{22}$ into the formula for the memory
function and using convolution approximation
for static vertices \cite{Goetze,SL} we get
\begin{widetext}
\begin{equation}\label{Mfinal}
M(k;z) = \frac{nD_0}{2}
\int \frac{d\mathbf{k}_1 d\mathbf{k}_2}{\left(2\pi\right)^3} 
\delta(\mathbf{k}-\mathbf{k}_1-\mathbf{k}_2)
\left[\hat{\mathbf{k}}\cdot
\left(c(k_1)\mathbf{k}_1+c(k_2)\mathbf{k}_2\right)\right]^2  
\frac{S(k_1) S(k_2)}
{z + \left[\frac{D_0 k_1^2/S(k_1)}
{1+LT\left(M(k_1;t)F(k_2;t)/S(k_2)\right)}
+ \left(1 \leftrightarrow 2 \right)\right]},
\end{equation}
\end{widetext}
where $n$ is the density and $c(k)$ is the direct correlation function.
Eqs. (\ref{Mkt}) and (\ref{Mfinal}) 
determine time dependence of density correlations 
and the memory function. 

Eqs. (\ref{Mkt}) and (\ref{Mfinal}) predict an ergodicity breaking
transition. In the non-ergodic regime $F(k;t)$ has a non-zero long-time
limit, $\lim_{t\rightarrow\infty} F(k;t) = f(k)S(k)$, where
$f(k)$ is called a non-ergodicity parameter. It follows from 
Eq. (\ref{Mkt}) that in this regime also the memory function has a non-zero
long-time limit, $\lim_{t\rightarrow\infty} M(k;t) = m(k)D_0k^2/S(k)$,
and that $f(k)$ and $m(k)$ are related by 
\begin{equation}\label{fm}
\frac{f(k)}{1-f(k)} = m(k).
\end{equation}
Using (\ref{Mfinal}-\ref{fm}) we get a self-consistent equation for 
$f(k)$:
\begin{widetext}
\begin{equation}\label{fnew}
\frac{f(k)}{1-f(k)} = 
\frac{n}{2k^2}
\int \frac{d\mathbf{k}_1 d\mathbf{k}_2}{\left(2\pi\right)^3} 
\delta(\mathbf{k}-\mathbf{k}_1-\mathbf{k}_2)
\left[\hat{\mathbf{k}}\cdot
\left(c(k_1)\mathbf{k}_1+c(k_2)\mathbf{k}_2\right)\right]^2  
\frac{S(k) S(k_1) S(k_2) f(k_1) f(k_2)}
{1 + (1-f(k_1))(1-f(k_2)) }.
\end{equation}
\end{widetext}
One should note that the right-hand-side of an
analogous self-consistent equation derived from MCT has
a similar form; the difference is that within MCT the right-hand-side is
a quadratic functional of $f(k)$ \cite{Goetze} whereas in the present
approach it 
includes terms of all orders in $f(k)$.

For low enough densities Eq. (\ref{fnew}) has only trivial solutions 
(\textit{i.e.} $f(k)=0$). For the hard-sphere interaction a non-trivial
solution appears at $n_g\pi\sigma^3/6=\phi_g=.549$. 
Qualitatively, the ergodicity breaking transition is similar to
that predicted by MCT: 
$f(k)$ has a jump at the transition.
Also, $f(k)$ at the transition is similar to that
of MCT at the MCT transition, 
$\phi_g^{MCT}=.525$ (Fig. 1). 

\begin{figure}
\resizebox{!}{5.8cm}{\includegraphics{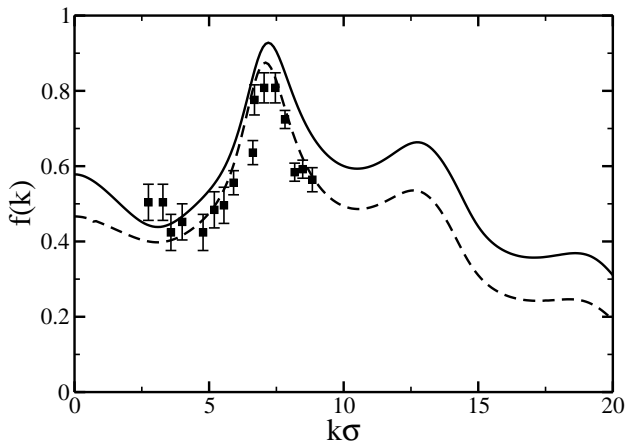}}
\caption{\label{fig1}Non-ergodicity parameter. Lines: theoretical
predictions at the ergodicity 
breaking transition; solid line: our theory, $\phi_g=.549$; dashed line:
MCT, $\phi_g^{MCT}=.525$. Symbols: experimental data taken at $\phi=.563$
\cite{vanM}.}
\end{figure}

The factorization approximation proposed here is the simplest possible
one. There are two 
ways to improve upon it. First, one could
try to include in an approximate way the off-diagonal blocks of $\mathcal{M}$.
To this end 
one could express them in terms of a triple-density correlation
function and then factorize this function into a product of three 
density correlation functions. Second, since the frequency matrix involves
only static correlations, one could try to include it in a more 
sophisticated way. For example, one could 
include two-particle dynamics exactly \cite{Arun}. 
The second extension 
could describe glassy dynamics in systems without static 
correlations \cite{SS}. 

To summarize, we proposed a new theory for dynamics of concentrated
suspensions and the colloidal glass transition. The theory goes beyond
MCT in that includes, in an approximate way, time-dependent pair-density
fluctuations. In contrast to an earlier approach\cite{Oppen}, the present
one uses pair-density 
correlation function evolving with one-particle irreducible dynamics.
The new theory predicts an ergodicity breaking transition similar
to that of MCT, but at a higher density.

The author benefited from inspiring discussions with 
Rolf Schilling and Arun Yethiraj; support by
NSF Grant No. CHE-0111152 is gratefully acknowledged.


\begin{thebibliography}{99}
\bibitem{review} For a general introduction see, \textit{e.g.},
J.K.G. Dhont, \textit{An Introduction to Dynamics of Colloids}
(Elsevier, New York, 1996); for a very recent overview emphasizing
connections between diffusional and viscoelastic properties see
G. N\"{a}gele, J. Phys. Cond. Matter \textbf{15}, S407 (2003).
\bibitem{expts} See, \emph{e.g.}, W.K. Kegel and A. van Blaaderen,
Science \textbf{287}, 290 (2000);
E.R. Weeks and D.A. Weitz, Phys. Rev. Lett. \textbf{89}, 095704 (2002).
\bibitem{colloglass} See, \textit{e.g.}, W. H\"{a}rtl,
Curr. Opin. Colloid Interface Sci. \textbf{6}, 479 (2001),
K.A. Dawson, \textit{ibid} \textbf{7}, 218 (2002).
\bibitem{Cates} M.E. Cates, cond-mat/0211066.
\bibitem{Goetze} W. G\"{o}tze, in \textit{Liquids, Freezing and Glass 
Transition}, J.P. Hansen, D. Levesque, and J. Zinn-Justin, eds. 
(North-Holland, Amsterdam, 1991).
\bibitem{SL} G. Szamel and H. L\"{o}wen, Phys. Rev. A \textbf{44}, 8215 (1991).
\bibitem{commentHI} This applies to systems in which
hydrodynamic interactions can be neglected: strongly charged suspensions
or simulated many-particle systems with Brownian dynamics. For real 
hard-sphere-like suspensions neglecting hydrodynamic 
interactions (or including them via a rescaling procedure
(M. Medina-Noyola, Phys. Rev. Lett. \textbf{60}, 2705 (1988))) 
constitutes an additional approximation; the importance of this approximation
is largely unknown. 
\bibitem{commentMCTder} There are several ways to derive MCT. Here 
we discuss the original projection operator technique \cite{Goetze} since it
is the one used to derive the new theory.
\bibitem{commentirr} Note that for Brownian systems one has to use
irreducible memory function 
(S.J. Pitts and H.C. Andersen, J. Chem. Phys. \textbf{113}, 3945 (2000); 
see also Ref. \cite{CHess}).
\bibitem{commentBvN} For Newtonian systems expressing the memory
function in terms of pair-density correlation
function neglects couplings to current modes. 
Within extended MCT currents 
restore ergodicity and cut off the ideal glass transition
(W. G\"{o}tze and L. Sj\"{o}gren, Z. Phys. B \textbf{65}, 415 (1987)). 
\bibitem{Naegele} G. N\"{a}gele and J. Bergenholtz, J. Chem. Phys.
\textbf{108}, 9893 (1998).
\bibitem{Fuchs} M. Fuchs and M.E. Cates, Phys. Rev. Lett. \textbf{89},
248304 (2002).
\bibitem{DawsonFuchs} K.A. Dawson, G. Foffi, M. Fuchs, \textit{et al.},
Phys. Rev. E \textbf{63}, 011401 (2001).
\bibitem{vanM} W. van Megen, S.M. Underwood, and P.N. Pusey,
Phys. Rev. Lett. \textbf{67}, 1586 (1991).
\bibitem{Kob} M. Nauroth and W. Kob, Phys. Rev. E \textbf{55}, 657 (1997);
this work discusses a Newtonian system; within MCT the location of the
transition does not depend on the microscopic dynamics \cite{SL}.
\bibitem{SS} R. Schilling and G. Szamel, Europhys. Lett.{\bf 61}, 207 (2003).
\bibitem{Oppen} C.Z.-W. Liu and I. Oppenheim, Physica A \textbf{247}, 183 
(1997).
\bibitem{Arun} For a similar idea in a different context see K. Miyazaki
and A. Yethiraj, J. Chem. Phys. \textbf{117}, 10448 (2002).
\bibitem{commentHI2} Following prior works on the colloidal glass
transition, hydrodynamic interactions are neglected.
\bibitem{CHess} B. Cichocki and W. Hess, Physica A \textbf{141}, 475 (1987).
\bibitem{comment2b} Eq. (\ref{projection}) is exact 
for systems with pairwise-additive interactions. 
\bibitem{GS} G. Szamel, to be published.
\end{thebibliography}
\end{document}